\documentclass[epj]{svjour}

\usepackage{graphicx}
\usepackage{amsmath,amsfonts,amssymb}

\begin{document}
\title{Continuous Markovian model for L\'evy random walks \\ with superdiffusive and superballistic regimes}
\titlerunning{Continuous Markovian model for L\'evy random walks}
\author{Ihor Lubashevsky\inst{1,2}\thanks{\,\email{ialub@fpl.gpi.ru}}
       \and
       Andreas Heuer\inst{3,5}\thanks{\, \email{andheuer@uni-muenster.de}}
       \and
       Rudolf Friedrich\inst{4,5}\thanks{\, \email{fiddir@uni-muenster.de}}
       \and
       Ramil Usmanov\inst{6}
}                     
\institute{
\mbox{University of Aizu, Tsuruga, Ikki-machi, Aizu-Wakamatsu City, Fukushima 965-8580, Japan}
    \and
\mbox{A.M. Prokhorov General Physics Institute, Russian Academy of Sciences, Vavilov Str. 38, Moscow 119991, Russia}
    \and
\mbox{Institut f\"ur Physikalische Chemie, Westf\"alische Wilhelms Universit\"at M\"unster,  Corrensstr. 30, 48149 M\"unster, Germany}
    \and
\mbox{Institut f\"ur Theoretische Physik, Westf\"alische Wilhelms Universit\"at M\"unster, Wilhelm-Klemm. 9, 48149 M\"unster, Germany}
    \and
\mbox{Center of Nonlinear Science CeNoS, Westf\"alische Wilhelms Universit\"at M\"unster, 48149 M\"unster, Germany}
    \and
\mbox{Moscow Technical University of Radioengineering, Electronics, and Automation, Vernadsky pr., 78, 117454 Moscow, Russia}
}
\date{Received: date / Revised version: date}
%
\abstract{
We consider a previously devised model describing L\'evy random walks (Phys. Rev E  \textbf{79}, 011110; \textbf{80}, 031148,
(2009)). It is demonstrated numerically that the given model describes L\'evy random walks with superdiffusive, ballistic, as well as superballistic dynamics. Previously only the superdiffusive regime has been analyzed. In this model the walker velocity is governed by a nonlinear Langevin equation. Analyzing the crossover from small to large time scales we find the time scales on which the velocity correlations decay and the walker motion essentially exhibits L\'evy statistics. Our analysis is based on the analysis of the geometric means of walker displacements and allows us to tackle probability density functions with power-law tails and, correspondingly, divergent moments.
} 
\maketitle

\section{Introduction}\label{Intr}

The L\'evy-type behavior of anomalous Brownian motion has been found in a
large variety of systems
in nature (see, e.g., \cite{Mandl,BG,Biol1,Zasl1,Klaft1}). It is characterized
by the scaling law
$|\delta \mathbf{r}|\propto t^\beta$ of the particle displacement $|\delta
\mathbf{r}|$
during the time interval $t$ with an exponent $\beta > 1/2$.
The cases of $1/2< \beta < 1$ and $\beta > 1$ are usually referred to as the
superdiffusive and superballistic regimes of the L\'evy-type dynamics, respectively.

The continuous time random walk model, including its decoupled and coupled
implementations,
is now a widely used approach to describe L\'evy-type stochastic processes
(see, \cite{Klaft1,CTRW2} and references therein). Within the classical formulation
L\'evy flights are modeled as a sequence of random independent steps whose
lengths are
distributed according to a power-law probability density with diverging
moments. Unfortunately, this model does not generate continuous trajectories.
Indeed, by one step a L\'evy walker can make a very long jump  and
such events in spite of being rather rare contribute substantially to the
walker displacements.
This feature is responsible for considerable problems met in describing
L\'evy-type
processes in heterogeneous media and systems with boundaries.
The coupled continuous time random walks \cite{CTRW2}, also referred to as
L\'evy random walks,
do admit a representation of the  walker motion in terms of continuous
trajectories; this model just
mimics the stochastic process at hand as a sequence of random independent
steps whose length and duration are, in turn, random variables. The latter
enables one to introduce a finite velocity of the walker motion during one
step by connecting its initial and terminal points with a straight line and
considering the walker motion along it uniform. Unfortunately, the resulting
stochastic process looses the Markov property in the strict sense.

There have been several attempts to develop a description of L\'evy-type
random walks starting from a certain Markovian continuous process with, may
be, nonlinear (multiplicative) Langevin forces. In particular, it is a
description of the decoupled continuous time random walks based on the
corresponding
Langevin equation \cite{Fogedby}, the combination of a conventional stochastic
process governed by a linear Langevin equation with additive noise where,
however,
time itself is a random nondecreasing variable \cite{HN}, models bases on
stochastic differential equations with non-Gaussian L\'evy white noise
\cite{NGLN1,NGLN2}, and
the semiclassical Fokker-Planck equation describing quantum interaction of
atoms with
laser beams \cite{F1,F2}, as well as models using the Boltzmann equation with
scattering characterized by power-law distribution \cite{Fr1,Fr2,Fr3}. There
is also an approach to describe
 anomalous diffusion based on the nonlinear Fokker-Planck equation
derived using the generalized entropy \cite{Tsalis}.

In systems far from thermodynamics equilibrium and complex systems  nonlinear
stochastic processes are widely met and have been studied in connection with
the appearance of power-law distributions and long-time tails of time
correlation functions
(see, e.g., \cite{K1,K2} as well as \cite{K3,KW} and references therein).

The generalized Cauchy stochastic process, which is governed by the following
Langevin equation with additive and multiplicative noise \cite{KW}
\begin{subequations}\label{int:1}
\begin{equation}\label{int:1a}
  \tau\frac{dv}{dt} = - [\lambda + \sqrt{2\tau}\xi_m(t)] * v  + \sqrt{2\tau} v_a \xi_a(t)
\end{equation}
has been employed to study stochastic behavior of various nonequilibrium
systems, in particular, lasers \cite{K1}, on-off intermittency \cite{Nak5},
economic activity \cite{Nak9}, passive scalar field advected by fluid
\cite{Nak11}, etc.
Here $v$ is a certain random variable, $\tau$ is a characteristic
``microscopic'' time scale
of its dynamics, $\lambda$ is a friction constant, $\xi_m(t)$ and $\xi_a(t)$
are independent
components of white noise of unit amplitude, the parameter $v_a$ quantifies
the intensity of additive noise, and the asterisk indicates that
equation~\eqref{int:1a}
is written in the H\"anggi-Klimontovich form.
The coefficients $\sqrt2$ are introduced into \eqref{int:1a} for the sake of
convenience.
Equation~\eqref{int:1a} is equivalent to the following Langevin equation
with a single white noise source $\xi(t)$ and the amplitude
depending nonlinearly on the random variable $v$ \cite{KW}
\begin{equation}\label{int:1b}
  \tau\frac{dv}{dt} = - \lambda v + \sqrt{2\tau\big(v_a^2+v^2\big)} \xi(t)
\end{equation}
\end{subequations}
because both of them are described by the same Fokker-Planck equation for the
probability density $P(v,t)$
\begin{equation}\label{int:2}
  \tau\frac{\partial P}{\partial t} = \frac{\partial}{\partial v}\left[
   \big(v_a^2+v^2\big)\frac{\partial P}{\partial v} + \lambda v P
  \right]\,.
\end{equation}
Appealing to qualitative arguments and a numerical example  Sakaguchi
\cite{Saka}
demonstrated  that the generalized Cauchy process~\eqref{int:1a} can give rise
to L\'evy flights in the space $x$ on time scales $t\gg \tau$ if the random
variable $v$ is
regarded as the velocity $v=dx/dt$ of a Brownian particle.

Previously \cite{we1,we2} we analyzed in detail such random walks governed by
the Langevin equation~\eqref{int:1b} when the induced L\'evy-type process is
superdiffusive. In particular, it has been demonstrated \cite{we1} that on
time scales $t\gg\tau$, related with the crossover to L\'evy-type behavior, the
scaling law $\delta x\propto t^\beta$ with the exponent $1/2<\beta<1$ is due
to the impact of extremely large fluctuations in the velocity pattern
$\{v(t)\}$. Roughly speaking, the peak velocity $\{v_\text{max}(t)\}$ times
the duration $\tau$, where $v_\text{max}$ is the maximal value of the velocity attained
during the time interval $t$, is the main contribution to the particle
displacement $\delta x(t)$. As a consequence,
the estimate $\delta x\sim v_\text{max}\tau$ holds and the characteristic scaling
law is related to the statistics of the occurrence of
extreme fluctuations in the observation interval $t$.

It should be noted that the constructed random process neither belongs  to the
decoupled nor to coupled continuous time random walks. Indeed, on one hand side,
this Brownian particle moves continuously and, on the
other hand, its motion with the ``elementary'' time fragment $t$ is not
uniform but extremely heterogeneous. The latter is actually the mechanism
preserving the Markov property of such random walks. Naturally on time scales
about $\tau$ and the corresponding spatial scales $\ell= v_a \tau$ random
walks governed by the Langevin equation~\eqref{int:1b} are ballistic,
$\delta x(t)\propto t$, because of the strong velocity correlations on such
scales. In our paper \cite{we2} we have developed a special singular
perturbation technique and proved, for the superdiffusive
regime, that the stochastic process~\eqref{int:1b} indeed describes the
L\'evy-type random walks on time scales $t\gg\tau$. At the critical value
$\lambda_c$ separating the superdiffusive and superballistic regimes the
constructed perturbation technique exhibits some singularities at intermediate
steps, but the final results turn out to be free from any singularity at
$\lambda = \lambda_c$. This problem seems to be related to the well-known conflict with dynamics in
constructing the superballistic L\'evy flights, extensively discussed in the review \cite{Zasl2}.

The purpose of the present paper is, first, to demonstrate that the developed model
\cite{we1,we2} not only describes the superdiffusive L\'evy-type random walks
but also superballistic ones. A second aim is to analyze the cross\-over between
the time scales $t\sim\tau$ and $t\gg\tau$ in the behavior of the walker
displacements. The latter result will clarify the behaviour at the time
scales for which velocity correlations become insignificant and
L\'evy-type dynamics arises.

It should be noted that superballistic transport is expected to arise
in such systems as atmospheric turbulence \cite{sb1,sb2}, two-dimensional
turbulence \cite{sb3}, a collisionless plasma driven in magnetic field
reversal with an electric field in the presence of magnetic turbulence
\cite{sb4}. It also has been observed experimentally in a fully developed
turbulent laboratory water flow \cite{sb5} as well as in a one-dimensional
expansion of cold atomic clouds in a magneto-optical trap under the impact of
laser beams \cite{sb6}.
In spite of these objects, strictly speaking, do not belong to the systems
under consideration,
it was found that the notion of L\'evy-type random walks \cite{sb2,sb3} or
stochastic differential equations \cite{sb6} gives rise to the adequate
description of these superballistic phenomena.
Large-scale animal movements in searching resource patches, mates, etc. or
dispersal also exhibit the L\'evy-type pattern (for a review see
\cite{For1,For2,Fog3})
and there are conditions when the optimal random search becomes of the
ballistic type as a limit case of L\'evy-type motion \cite{Fog4,Fog5}.
L\'evy type behavior is relevant to stock indices, too, and the indices
matching the superballistic L\'evy flights were found, in particular,
for the Brazil BVSP index and the Argentine MERV index \cite{EPh1},
for foreign exchange rate in Brazil and South Korea \cite{EPh2a,EPh2b}, and
the absolute value of the returns in the Indian market indices \cite{EPh3}

\section{Model}\label{sec:2}
Let us restrict our consideration to a dimensionless description of the
stochastic processes at hand and, thereby, measure time in units of $\tau$,
i.e. $t\to\tau t$ and spatial variables in units of $\ell$, i.e. $x_i \to \ell
x_i$. In these units the continuous Markovian random walks under consideration
are governed by the following equations

\begin{align}
\label{mod:1}
    \frac{dx_i}{dt} & = v_i\,,
\\
\label{mod:2}
    \frac{dv_i}{dt} & = -\left(n+\alpha\right)v_i +
\sqrt{2(1+\mathbf{v}^2)}*\xi_i(t)\,.
\end{align}
Here $\mathbf{r}=\{x_1,\dots,x_n\}$ is the current position of a random walker
in the Euclidean $n$-dimensional space $\mathbb{R}^n$ and
$\mathbf{v}=\{v_1,\dots,v_n\}$
is the velocity of its motion in this phase space,
$\{\xi_1(t),\ldots\xi_n(t)\}$
is the collection of mutually independent components of Gaussian white noise,
\begin{equation}\label{mod:3}
    \left<\xi_i(t)\right> = 0\,,\quad\text{and}\quad
\left<\xi_i(t)\xi_{i'}(t')\right> = \delta_{ii'}\delta(t-t')\,,
\end{equation}
and, as before, the asterisk indicates that equation~\eqref{mod:2} is written
in the H\"anggi-Klimontovich form. The parameter $\alpha$ is assumed to be
taken from the interval
\begin{equation}\label{mod:4}
 0 < \alpha < 2\,.
\end{equation}

Previously \cite{we2} dealing with this model for $1<\alpha<2$ we demonstrated that in the limit $t\gg1$ the characteristic amplitude $|\delta\mathbf{r}|$ of the walker displacements scales with time $t$ as  $|\delta\mathbf{r}(t)| \propto t^{1/\alpha}$, matching the superdiffusive L\'evy random walks. Besides, the probability density $P(\mathbf{\delta r},t)$ of the walker displacement $\delta\mathbf{r} \in \mathbb{R}^n$ is found to possess the required asymptotics $P(\mathbf{\delta r},t)\propto |\delta\mathbf{r}|^{-(n+\alpha)}$ when $|\delta\mathbf{r}|\to\infty$.  These expressions stem directly from the obtained dependence of the generating function for $t\gg1$
\begin{equation}\label{mod:gf}
    G(\mathbf{k},t) = \left<e^{i\mathbf{kr}}\right> =
\exp\left\{-\Lambda_\text{min}t|\mathbf{k}|^\alpha\right\},
\end{equation}
where $\Lambda_\text{min} > 0$ is the minimal eigenvalue of the corresponding Fokker-Planck equation. In the 1D-case it is given by the expression
\begin{equation}\label{mod:Lambda}
    \Lambda_\text{min} = \frac{\Gamma\left(\dfrac{2-\alpha}{2}\right)}
    {2^{\alpha-1}\Gamma\left(\dfrac{\alpha}{2}\right)\Gamma\left(\alpha\right)}\,,
\end{equation}
where $\Gamma(\dots)$ is the gamma function.

As noted before all these final results do not exhibit singularities when
$\alpha\to 1$ whereas the constructed perturbation technique does at its
intermediate steps. This feature poses a question as to whether the
stated model~\eqref{mod:1}, \eqref{mod:2} does describe the L\'evy random
walks in a wider region of parameters including the whole
interval~\eqref{mod:4} and comprising superdiffusive as well as superballistic dynamics.

In the next section this hypothesis will be justified numerically; here,
we present some heuristic arguments for the absence of any singularity in the
behavior of the given system when the parameter $\alpha$ is taken from the interval $0< \alpha \leq 1$.

\subsection*{Canonical form of equation~\eqref{mod:2}}

Dealing with the one dimensional stochastic processes governed by the Langevin
equation~\eqref{mod:2} for $n=1$ let us make use of the nonlinear transformation \cite{KW}
\begin{equation}\label{mod:5}
    v = \sinh(u)
\end{equation}
reducing equation~\eqref{mod:2} to the following Langevin equation with additive noise
\begin{equation}\label{mod:6}
  \frac{du}{dt} = -\alpha \tanh(u) + \sqrt{2}\xi(t)\,.
\end{equation}
In obtaining expression~\eqref{mod:6} we, first, converted
equation~\eqref{mod:2} into the Stratonovich form and only then performed the
transformation~\eqref{mod:5}. According to Konno and Watanabe \cite{KW}
expression~\eqref{mod:6} can be regarded as the canonical form of such
stochastic differential equations with multiplicative noise.

From the Fokker-Planck equation 
\begin{equation*}
\frac{\partial P}{\partial t} = \frac{\partial }{\partial u}\left[
    \frac{\partial P}{\partial u} + \alpha\tanh(u) P
\right]
\end{equation*}
corresponding to
the Langevin equation~\eqref{mod:6} we obtain the steady state distribution of the
random variable $u$
\begin{equation}\label{mod:7}
  P_c(u) = \frac{A_c}{[\cosh(u)]^\alpha}
   \,,
\end{equation}
where $A_c$ is the normalization constant. In particular, for $\alpha \sim 1$
the constant $A_c\sim 1$ too and for $\alpha \ll 1$ the estimate $A_c =\alpha/2$ holds.
From the Langevin equation~\eqref{mod:6} we can estimate
the time scale $T_c$ characterizing the formation of the steady state
distribution~\eqref{mod:7}
as $T_c\sim 1$ for $\alpha \sim 1$ and $T_c\sim 1/\alpha^2$ for $\alpha \ll 1$.

We conclude that the statistical properties of the random variable $u$ cannot exhibit any singularity for the parameter $\alpha$ lying inside the interval under consideration, i.e.  $\alpha\in(0,2)$. The parameter $\alpha$ just affects the weight of extreme fluctuations with respect to their contribution to the walker displacement, changing the exponent of its time dependence. Moreover, in studying these extreme fluctuations we can approximate the function $\tanh(u)$ by the signum function $\text{sign}(x)$, i.e., make use of the replacement $\tanh(u)\to\text{sign}(x)$ and perform the transformations $u\to u/\alpha$ and $t\to t/\alpha^2$. This reduces equation~\eqref{mod:6} to the following equation
\begin{equation}\label{mod:6a}
  \frac{du}{dt} = -\text{sign}(u) + \sqrt{2}\xi(t)\,.
\end{equation}
In deriving this equation the white noise transformation $\xi(t)\to \alpha\xi(t)$ caused by the given time renormalization and stemming from the induced transformation of the correlation function
\begin{equation*}
    \left<\xi(t)\xi(t')\right> = \delta(t-t') \underset{t\to t/\alpha^2}{\longrightarrow} \alpha^2\delta(t-t')
\end{equation*}
has been taken into account. Equation~\eqref{mod:6a} does not contain any parameter, which seems to be the reason for
the universal behavior of such systems.

Therefore, it is quite natural to expect that for $0<\alpha\leq 1$
model~\eqref{mod:1}, \eqref{mod:2} also describes the L\'evy-type random walks
with, maybe, the superballistic dynamics provided the expression
$\delta x(t)\propto t^{1/\alpha}$ holds. The singularities arising at
intermediate steps of the perturbation technique seem to be caused
by anomalous behavior of integrals similar to
\begin{equation*}
   \int\limits_{-\infty}^{\infty} P(x,t) \sin(kx)\, dx \propto
   \int\limits_{-\infty}^{\infty} \frac{\sin(kx)}{|x|^{1+\alpha}}\, dx
\end{equation*}
as $k\to0$. For $\alpha < 1$ this integral also leads
to nonanalytic dependencies on the wave number $k$ and must vanish for the symmetry reasons only.

\section{Numerical results}\label{sec:4}

In order to justify the fact that the model~\eqref{mod:1}, \eqref{mod:2}
does describe the L\'evy-type random walks on time scales
$t\gg1$ for $0<\alpha<2$ the system of equations~\eqref{mod:1},
\eqref{mod:2} has been solved numerically. Then the obtained results
were used to calculate the geometric mean value
\begin{equation}\label{sec4:1}
    \ln\!\big[\overline{r}_g(t)\big] = \left<\ln\!\big(|\delta\mathbf{r}|\big)\right>
\end{equation}
of the walker's displacement  $\delta \mathbf{r}$ during the time interval $t$ as well as the probability density function $P(|\delta\mathbf{r}|,t)$. It should be pointed out that logarithmic mean values similar to \eqref{sec4:1} form a rather efficient tool for analyzing statistical data with power-law distributions possessing diverging moments \cite{KW}. It is due to the geometric mean converging for any power-law distribution that can be normalized to unity and, thus, enables us to analyze superdiffusive and superballistic regimes with any possible value of the exponent $\alpha\in(0,2)$ using the same quantities. The geometric mean value introduced via expression~\eqref{sec4:1} can be also treated as the limit case of constructing the Hurst exponent $H(\varepsilon)$
\begin{equation*}
  \left<|\delta \mathbf{r}|^\varepsilon\right>^{1/\varepsilon} \propto t^{H(\epsilon)} \quad\text{for}\quad\varepsilon\to+0\,.
\end{equation*}
In addition, when the random variables $\delta\mathbf{r}(t)$ are characterized by an autonomous probability density $\mathcal{P}\left(|\delta\mathbf{r}|/t^{\beta}\right)$ expression~\eqref{sec4:1} reads
\begin{equation*}
    \ln\!\big[\overline{r}_g(t)\big] = \int\limits_0^\infty \ln(\zeta)\mathcal{P}(\zeta)d\zeta + \beta \ln(t)\,,
\end{equation*}
which allows us to separate the time dependence of the scaling low $|\delta\mathbf{r}(t)|\propto t^\beta$ from the corresponding numerical coefficients.

If the generating
function $G(\mathbf{k},t)$ for the quantities $\{\delta \mathbf{r}\}_t$
is given by expression~\eqref{mod:gf} formula~\eqref{sec4:1} yields
\begin{equation}\label{sec4:2}
    \overline{r}_g(t) = \Lambda_g\cdot t^{1/\alpha},
\end{equation}
as shown in Appendix~\ref{app1}. In particular, we obtain
in the 1D-case the coefficient
\begin{align}
\label{sec4:3a}
    \Lambda_g|_{n=1} &= \Lambda_\text{min}^{1/\alpha}\cdot\exp\left\{\gamma\, \frac{(1-\alpha)}{\alpha}\right\},
\\
\intertext{and for the 2D-case}
\label{sec4:3b}
    \Lambda_g|_{n=2} &= 2\Lambda_\text{min}^{1/\alpha}\cdot\exp\left\{\gamma\, \frac{(1-\alpha)}{\alpha}\right\}.
\end{align}
Here $\gamma\approx  0.5772$ is the Euler-Mascheroni constant.

If the random variable $\delta\mathbf{r}$ obeys the L\'evy statistics,
then the asymptotics of the probability density $P(|\delta\mathbf{r}|,t)$
is also determined by equation~\eqref{mod:gf} for the generating
function $G(\mathbf{k},t)$. Moreover, provided the walker displacement
$\delta\mathbf{r}$ is measured in units of the corresponding
geometric mean $\overline{r}_g(t)$, i.e. in terms of
$\rho=|\delta\mathbf{r}|/\overline{r}_g(t)$, its probability
density is characterized by the asymptotics (Appendix~\ref{app2})
\begin{equation}\label{sec4:4}
    \mathcal{P}(\rho) = \frac{\varpi_{\rho}}{\rho^{1+\alpha}}\qquad\text{as}\quad\rho\to\infty\,,
\end{equation}
where in the 1D-case the coefficient reads
\begin{align}
\label{sec4:5a}
    \varpi_{\alpha}|_{n=1} & = \frac2{\pi} \sin\left(\frac{\pi\alpha}2\right) \Gamma\left(1+\alpha\right)e^{\gamma(\alpha-1)}
\\
\intertext{whereas in the 2D-case}
\label{sec4:5b}
    \varpi_{\alpha}|_{n=2} &= \frac2{\pi} \sin\left(\frac{\pi\alpha}2\right) \left[\Gamma\left(1+\frac\alpha2\right)\right]^2e^{\gamma(\alpha-1)}\,.
\end{align}
Expressions~\eqref{sec4:2} and \eqref{sec4:4} are the key results
for the numerical verification of the L\'evy-type dynamics of the
random walks under consideration. Namely, analyzing the numerical
simulation data it is possible to verify whether the constructed
time dependence of the geometric mean $\overline{r}_g(t)$ and
the probability density $P(\mathbf{r},t)$ actually do obey
expressions~\eqref{sec4:2} and \eqref{sec4:4}.

\begin{figure*}
\begin{center}
\includegraphics[width=\textwidth]{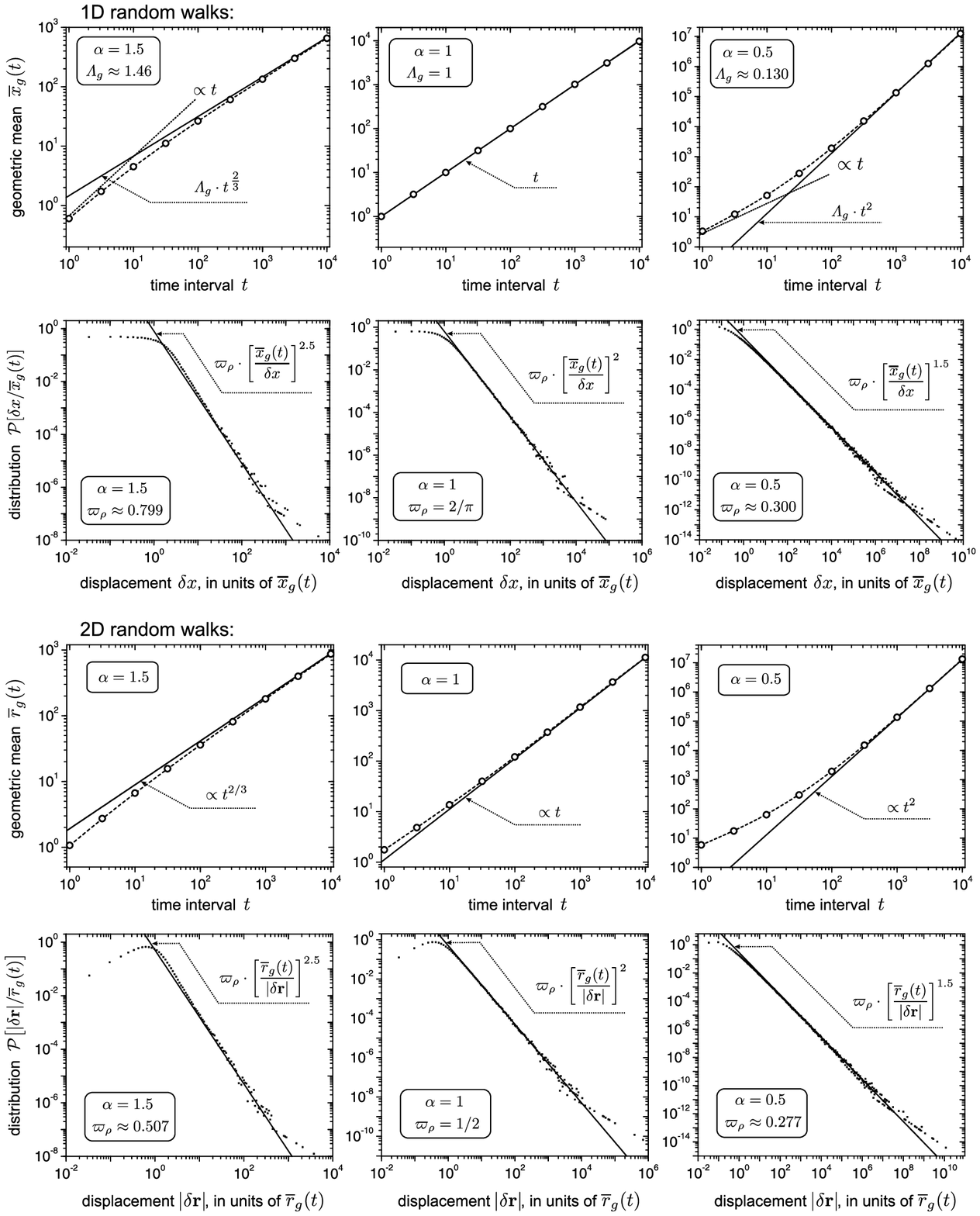}
\end{center}
\caption{The geometric mean of the walker displacements versus
the observation time interval and the probability density of the
walker displacements. In simulating 1D and 2D models the integration time
steps was set equal to $dt = 0.05$ and $dt = 0.02$, respectively, the
integration time was $T = 10^8$. The other parameters are shown in the panels.}
\label{Fig1}
\end{figure*}

For the 1D-case the system of equations~\eqref{mod:1}, \eqref{mod:2} has been
solved numerically employing the stochastic Runge-Kutta method SRI2W1
approximating one-dimensional Wiener process It\^o equations with strong
order (4.0,1.5) \cite{SRK1}. The time step $dt = 0.05$ for
integration yields results, which were stable with respect to a decrease
of $dt$.
The integration time was $T = 10^8$. For the 2D-case this system was
solved using the stochastic Runge-Kutta method RI1 approximating
multidimensional Wiener process It\^o equations with weak order (3,2)
\cite{SRK2}. The integration time step was chosen as $dt = 0.02$ yielding
stable results. The integration time was also $T = 10^8$.

Figure~\ref{Fig1} depicts the obtained results for the 1D and 2D models.
The relevant analytic time dependence of the geometric mean
$\overline{x}_g(t)$, in the 1D-case specified by
expressions~\eqref{mod:Lambda}, \eqref{sec4:2}, and \eqref{sec4:3a},
is also exhibited in Fig.~\ref{Fig1}. In particular,
for $\alpha = 1.5$, 1.0 and 0.5 we obtain $\Lambda_g \approx 1.46$,
$1$, and $0.130$, respectively. As seen in Fig.~\ref{Fig1} the
corresponding analytic dependence fits the numerical data quite well
if the time interval $t$ exceeds a value of about $t\sim100$. For time
scale $t\sim 1$ at which the velocity correlations should be significant
and, consequently, the walker behaviour has to be ballistic,
the effective exponent of the curve $\overline{x}_g(t)$ is about unity
as expected. Thus, the constructed curves actually describe the
crossover from the ballistic motion on time scales $t\sim1$
to the superdiffusive or superballistic regimes of the L\'evy-type
random walks for $t\gtrsim100$ for all the analyzed values of the
parameter $\alpha$. In the 2D-case an analytic expression for
the minimal eigenvalue $\Lambda_\text{min}$ is not known. Therefore, we can
compare only the slope of curves (in log-log scale) evaluated
numerically and the corresponding exponents of the theoretical
dependence~\eqref{sec4:2}. As seen in Fig.~\ref{Fig1} these values
again fit each other when $t\gtrsim100$ for the parameter
$\alpha \leq 1$ as well as $\alpha > 1$.

The numerical data for the probability density also exhibit the desired
asymptotics for both the 1D- and 2D-cases. According to
expression~\eqref{sec4:5a} in the 1D-case the coefficient $\varpi_\rho$
takes the values $\varpi_\rho\approx 0.799$, 0.300 for $\alpha = 1.5$, 0.5,
respectively, and $\varpi_\rho = 2/\pi$ for $\alpha = 1$.
For the 2D-case, by virtue of \eqref{sec4:5b}, it is 0.507, 0.277 and 1/2,
respectively. The corresponding theoretical asymptotics of the
probability density $\mathcal{P}(\rho)$ are also shown in Fig.~\ref{Fig1}.
In particular, as seen in this figure the function $\mathcal{P}(\rho)$
tends to asymptotics~\eqref{sec4:4} from above for $\alpha = 1.5$.

For the parameter $\alpha = 1.5$ the accuracy between the analytical and
numerical results is expected and can be regarded as a
verification of the used implementation of the employed numerical
algorithms. At the same time, for $\alpha = 1.0$ and 0.5 it is this
accuracy that justifies the L\'evy-type behavior of the stochastic
process governed by model~\eqref{mod:1}, \eqref{mod:2} for the
parameter $\alpha$ lying in the interval $0< \alpha\leq 1$.

It is necessary to note that the similarity of the plots for the 1D and 2D-cases with respect to the corresponding exponents is a consequence of the fact that the friction coefficient $(\alpha + n)$ entering equation~\eqref{mod:2} depends on the space dimension $n$. So in the  2D-case the components $\xi_1(t)$ and $\xi_2(t)$ of white noise individually undergo stronger dissipation and only their interplay leads to the same scaling law of the particle displacement magnitudes $\{|\delta\mathbf{r}|_t\}$. The difference between the plots showing the time dependence of the geometric means $\overline{x}_g(t)$ and $\overline{r}_g(t)$ for $\alpha=1$ enables us to claim that the ballistic-like behavior of the geometric mean on time scales $t\lesssim 1$ and its ballistic regime for $\alpha = 1$ must be characterized by the same exponent but not the numerical cofactors. So their visible equality in the 1D-case seems to be just coincidental. It also should be noted that in the 2D-case the probability density $\mathcal{P}(\rho)$ attains its maximum at a certain point $\rho_\text{max} >0$, whereas in the 1D-case this maximum is located at the origin $\rho = 0$. In fact, even in the 2D-case the probability density $P(\delta\mathbf{r},t) = P(|\delta\mathbf{r}|,t)$ of the \textit{random vector} $\delta\mathbf{r}$ gets its maximum at the origin too. However, the depicted probability density $\mathcal{P}(\rho)$ is proportional to the function $P(|\delta\mathbf{r}|,t)$ with a coefficient depending on $|\delta\mathbf{r}|$, namely, $\mathcal{P}(\rho)\propto|\delta\mathbf{r}|P(|\delta\mathbf{r}|,t)$, which causes the shift of its maximum from the origin.

\begin{figure}
\begin{center}
\includegraphics[width=0.85\columnwidth]{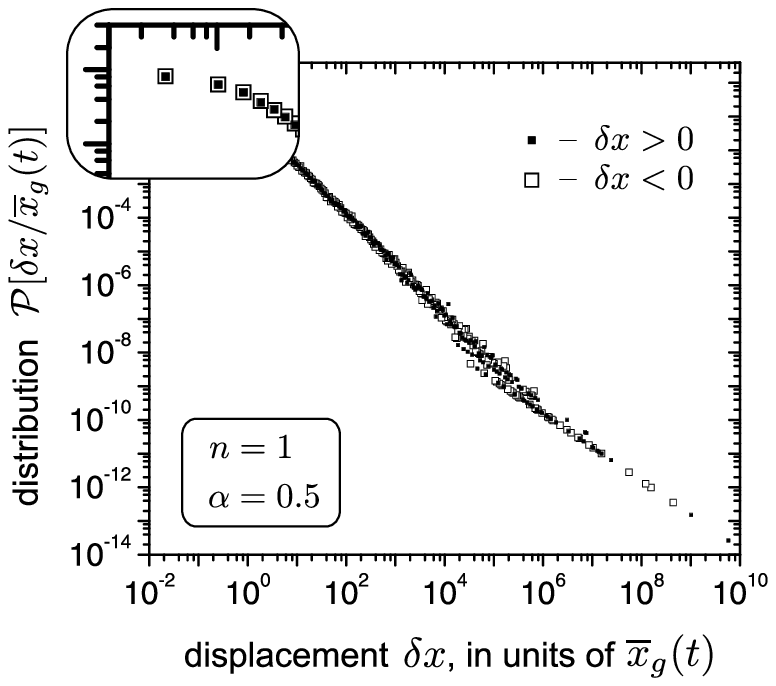}
\end{center}
\caption{The two side probability density of the walker displacements.
The data for the 1D-model with the parameter $\alpha=0.5$ matching the
superballistic regime are presented. The integration step was set
equal to $dt= 0.05$ and the integration time was $T = 10^8$.}
\label{Fig2}
\end{figure}

Finalizing the present section let us discuss ergodic properties of the
investigated stochastic process. Ergodicity usually means
that the ensemble average
\begin{equation*}
    \left<A\right>: = \int A P(A)dA\,,
\end{equation*}
where $P(A)$ is the probability density, and the time average
\begin{equation*}
\overline{A} := \frac{1}{t}\int\limits_0^{t}A(t')dt'
\end{equation*}
of an observable $A$  are equal in the
infinite-time limit $t\to\infty$. The condition
\begin{equation*}
 \left< \left(\overline{A} - \left<A\right>\right)^2\right>\to 0\quad \text{as}\quad t\to\infty
\end{equation*}
is widely used as a criterion for this equality \cite{erg1}. A system
obeying this criterion is said to be ergodic in the mean square sense.
Therefore, the notion of ergodicity also includes  the measure of
convergence. This type of ergodicity can be broken when the parameter
$A$ is distributed with a probability density possessing
power-law tails which cause the divergence of its moments
\cite{F2,erg2,erg3}.
Moreover, studying the ergodicity one may focus not on a single
trajectory of the system motion as a whole but treat it as a
collection of various fragments. As a result, a system could be regarded
as ergodic with respect to one variable and loose the ergodicity
according to the properties of another variable. Such a partial ergodicity is
met in the system at hand. In fact, each of the plots in Fig.~\ref{Fig1}
has been built using numerical data generated during the system motion
along a single sufficiently long trajectory. So the system ergodicity
with respect to quantities such as the geometric mean of walker
displacements and their distribution function is justified by the
accuracy between the presented numerical data and the theoretical
results related to the corresponding ensemble averages. As
illustrative example, Figure~\ref{Fig2} depicts the two-sided
probability density $P(\delta x,t)$ of one-dimensional random walks
for the parameter $\alpha = 0.5$, i.e. for the superballistic regime.
This probability density, in contrast to the preceding constructions,
takes into account also the direction of the walker displacements.
As seen in this figure even in the case of the superballistic regime
the probability density of the walker displacements constructed
by averaging over time is highly symmetric. In contrast, under
the same conditions such ergodicity does not hold with respect to any
quantity depending on the sign of the walker position at the $x$-axis.
Indeed, the mean time $t_a$ spent by the random walker inside a small
neighborhood $\mathbb{Q}_a$ of its initial point $x_0=0$ whose size is about $a$
can be estimated as
\begin{equation}\label{erg:1}
  t_a \sim \int\limits_0^\infty \frac{a}{\overline{x}_g(t)}\,dt\,.
\end{equation}
Estimate~\eqref{erg:1} stems directly from the expression for the walker residence time $t_a$ inside the domain $\mathbb{Q}_a$
\begin{equation}
  t_a = \int\limits_0^\infty dt\int\limits_{\mathbb{Q}_a} dx\, \delta[x-x(t)]
\end{equation}
for a given realization $\{x(t)\}$ of the walker trajectory and the estimate of the probability density of finding the walker at the initial point $x=0$
\begin{equation*}
    \mathcal{G}(0,t) = \left<\delta[x-x(t)]\right>|_{x=0} \sim \frac{1}{\overline{x}_g(t)}\,.
\end{equation*}
Integral~\eqref{erg:1} converges at the upper boundary for the parameter
$\alpha < 1$ since in this case the geometric mean
$\overline{x}_g(t)\propto t^{1/\alpha}$ grows faster than $t$.
As a result, there is a certain probability that the walker never returns
to its initial point and, thus, the walker will  practically spend
all the time in the region $x> 0$ or $x<0$ as $t\to\infty$.
As the averages over ensemble are concerned, the properties of
the given random walks should be symmetrical with respect to
the reflection $x\mapsto -x$.

\section{Conclusion}

We have considered the stochastic process governed by
equations~\eqref{mod:1}, \eqref{mod:2} and numerically
demonstrated that on time scales $t\gg1$ this process describes
the L\'evy-type random walks with the superballistic, ballistic,
and superdiffusive dynamics, depending on the value of the parameter
$\alpha$, namely, for $0<\alpha <1$, $\alpha = 1$, and $1<\alpha<2$,
respectively. In this way we have justified that the given model
is a continuous Markovian implementation of the L\'evy flights or
the L\'evy walks of all the main types that at the ``microscopic''
level admits the notion of continuous trajectories.

To analyze the statistical properties of stochastic processes for
which the moments of the probability density can diverge the notion
of geometric mean was used and the walker displacements were measured
in units of the corresponding geometric means.
This enabled us to compare  the numerical data and theoretical results
not only for one-dimensional systems but also for two-dimensional models
exemplifying the validity of the models for the case of
multi-dimensional random walks.

The analyzed crossover between the time scales $t\sim 1$ and
$t\gg1$ has shown that the L\'evy-type behavior arises on scales
$t\gtrsim100$. On scales $t\sim1$, where the velocity correlations
are significant, the random walks are ballistic. The direct simulation
of the system motion along individual sufficiently long
trajectories has demonstrated the fact that the system at hand is
partly ergodic, i.e. it is ergodic with respect to quantities such
as the geometric mean of the walker displacements and their distribution,
independently of whether the regime being superdiffusive or superballistic.
In the latter case, however, there are quantities for which
ergodicity is broken. At least the superballistic L\'evy-type random
walks described by the given model are only partly ergodic.

\begin{acknowledgement}
The authors are grateful to Hidetoshi Konno for his helpful comments and appreciate the financial support of the transregional research collaboration TRR 61  and the University of M\"unster as well as the partial support of the RFBR Grant 09-01-00736.
\end{acknowledgement}

\appendix
\section{Geometric mean and generating function}\label{app1}

Let us consider a random vector $\mathbf{r}$ of the
$n$-dimensional space $\mathbb{R}^n$ ($n = 1,2,\ldots$)
whose statistical properties are described by a given probability
density $P(\mathbf{r})$. Using this function we, first, construct
the generating function $G(\mathbf{k})$
\begin{equation}\label{app.0}
    G(\mathbf{k}):= \left<e^{i\mathbf{kr}}\right> = \int\limits_{\mathbb{R}^n}
e^{i\mathbf{kr}} P(\mathbf{r})\,d\mathbf{r}\,,
    \quad \mathbf{k}\in \mathbb{R}^n
\end{equation}
which is the Fourier transform of the probability density
and, then, the function
\begin{equation}\label{app.1}
    G_s(\mathbf{k}) := \frac1{\Omega_{n,s}}\int\limits_{\mathbb{R}^n}
    \frac{G(\mathbf{k}')}{|\mathbf{k}-\mathbf{k}'|^{n-s}}\,d\mathbf{k}'\,,
\end{equation}
where the exponent $0<s<1$, the coefficient
\begin{equation}\label{app.Omega}
    \Omega_{n,s} = \pi^{\tfrac{n}{2}}
    \frac{2^s\Gamma\left(\dfrac{s}{2}\right)}{\Gamma\left(\dfrac{n-s}{2}\right)}
\end{equation}
and, as before, $\Gamma(\ldots)$ is the gamma function.
The substitution of \eqref{app.0} into \eqref{app.1} yields the equality
\begin{equation}\label{app.2}
    G_s(\mathbf{k}) =  \int\limits_{\mathbb{R}^n} e^{i\mathbf{kr}}
P(\mathbf{r})\,d\mathbf{r}
    \left[
    \frac1{\Omega_{n,s}}\int\limits_{\mathbb{R}^n}
    \frac{e^{i\mathbf{qr}}}{|\mathbf{q}|^{n-s}}\,d\mathbf{q}
    \right]\,.
\end{equation}
A direct calculation of the last integral leads us to the expression
\begin{equation}\label{app.2a}
    G_s(\mathbf{k}) =  \int\limits_{\mathbb{R}^n}
\frac{e^{i\mathbf{kr}}}{|\mathbf{r}|^s} P(\mathbf{r})\,d\mathbf{r}\,.
\end{equation}
So, by definition, the mean value of the moment $1/|\mathbf{r}|^{s}$ is
\begin{equation}\label{app.3}
    \left<\frac1{|\mathbf{r}|^{s}}\right> :=
\int\limits_{\mathbb{R}^n}\frac{P(\mathbf{r})}{|\mathbf{r}|^{s}} \,d\mathbf{r}
  = \frac1{\Omega_{n,s}}\int\limits_{\mathbb{R}^n}
\frac{G(\mathbf{k})}{|\mathbf{k}|^{n-s}}\,d\mathbf{k}
\end{equation}
by virtue of \eqref{app.1} and \eqref{app.2a}.

Let us introduce the symmetrized generating function $g(k)$ by integrating the generating function $G(\mathbf{k})$ over the $n$-dimensional sphere $\mathbb{S}_{n,k} = \{\mathbf{k}:|\mathbf{k}|=k\}$
\begin{equation}\label{app.symm}
    g(k) :=  \frac1{S_n k^{n-1}} \oint\limits_{\mathbb{S}_k} G(\mathbf{k})\,
d\sigma\,,
\end{equation}
where $S_n k^{n-1}$ is the surface area of the sphere $\mathbb{S}_{n,k}$  and the coefficient
\begin{equation}\label{app.symmS}
    S_n = \frac{2\pi^{\tfrac{n}{2}}}{\Gamma\left(\dfrac{n}{2}\right)}\,.
\end{equation}
In particular, for the random vector $\mathbf{r}$ with the spherically symmetric distribution density $P(r)$ and, thus, the spherically symmetric generating function $G(k)$ the identity $g(k) \equiv G(k)$ holds. In these terms expression~\eqref{app.3} reads
\begin{multline}\label{app.4}
    \left<\frac1{|\mathbf{r}|^{s}}\right>
    = \frac{S_n}{\Omega_{n,s}}\int\limits_{0}^{\infty} k^{s-1}g(k)\, d k
\\
    {} =  -\frac{S_n}{s\cdot\Omega_{n,s}}\int\limits_{0}^{\infty}
k^{s}\frac{dg(k)}{dk}\, d k \,.
\end{multline}
Then using the series
\begin{align*}
    |\mathbf{r}|^{-s} & = 1 - s\ln|\mathbf{r}| + O\left(s^2\right)\,, \\
    k^s      & = 1 + s\ln k  + O\left(s^2\right)\,, \\
    2^s      & = 1 + s\ln 2  + O\left(s^2\right)\,, \\
    \Gamma\left(\frac{s}2\right) & = \frac1s\left[1 -\gamma\frac{s}2   +
O\left(s^2\right)\right]\,,\\
    \Gamma\left(\frac{n-s}{2}\right) & =
\Gamma\left(\frac{n}{2}\right)\left[1-\Psi\!\left(\frac{n}{2}\right)\frac{s}
2\right] + O(s^2)
\end{align*}
where, $\Psi(\ldots)$ is the digamma function and, as before, $\gamma \approx 0.5772$ is the Euler-Mascheroni constant, we expand the left and right sides of expression~\eqref{app.4} into series of $s$ up to the first order in $s$. In this way for the geometric mean value $\overline{r}_\text{g}$ of the random vector $\mathbf{r}$ defined as
\begin{equation}\label{app.gmv}
 \ln\left[\overline{r}_\text{g}\right] := \left<\ln|\mathbf{r}|\right>
\end{equation}
we obtain the formula
\begin{equation}\label{app.5}
 \ln\left[\overline{r}_\text{g}\right] =
 \int\limits_{0}^{\infty} \ln(k)\, \frac{dg(k)}{dk}\,dk - \gamma
 +  \frac12\left[\Psi\!\left(\frac{n}{2}\right) -
\Psi\!\left(\frac1{2}\right)\right]
\end{equation}
relating the geometric mean value $\overline{r}_\text{g}$ and the generating function $G(\mathbf{k})$ via $g(k)$. In deriving expression~\eqref{app.5} the equality
\begin{equation*}
    \Psi\left(\frac12\right) = -\gamma - 2\ln2
\end{equation*}
has been also taken into account. In particular, in the 1D-case the last summand is absent and for the 2D-case where $\Psi(n/2) = \Psi(1) = -\gamma$ its value is $[\Psi(1) - \Psi(1/2)]/2 = \ln2$. It should be noted that expression~\eqref{app.5} can be also regarded as a specific implementation of the Hadamard type fractional calculus (see, e.g., \cite{Kilbas}).

For the generating function $G(\mathbf{k}) = \exp\left(-\omega |\mathbf{k}|^\alpha\right)$, where $\omega>0$ and $\alpha >0$ are some positive constants, expression~\eqref{app.5} reads
\begin{equation}\label{app.6}
    \ln\overline{r}_\text{g} = \frac1{\alpha}\ln \omega +
\gamma\frac{(1-\alpha)}{\alpha}+ \frac12\left[\Psi\!\left(\frac{n}{2}\right) -
\Psi\!\left(\frac1{2}\right)\right].
\end{equation}
In obtaining the latter formula we have taken into account the equality
\begin{equation*}
    \int\limits_{0}^{\infty} \ln(k)\,e^{-k}\,dk = -\gamma\,.
\end{equation*}
A similar expression can be obtained for the $n$-dimensional generalized Cauchy distribution \cite{KonoPrivate}.

Formula~\eqref{app.6} immediately leads us to the desired expression~\eqref{sec4:2} for $\omega = \Lambda_\text{min} t$.

\section{Asymptotics of the probability density and generating
function}\label{app2}

As in Appendix~\ref{app1}, let us consider a random vector $\mathbf{r}\in \mathbb{R}^n$ whose statistical properties are specified by a given distribution density $P(\mathbf{r})$. Now, however, some additional assumptions about it are adopted. Namely, the leading term in the asymptotic series of the probability density $P(\mathbf{r})$ as $\mathbf{r}\to\infty$  is considered to be a spherically symmetrical power-law function. In other words, the equality
\begin{equation}
\label{appb:1}
    P(\mathbf{r})  =   \frac{\varpi}{r^{n+\alpha}} +
o\left(\frac1{r^{n+\alpha}}\right),
\end{equation}
where  $r = |\mathbf{r}|$ and the parameters $\varpi>0 $ and $0<\alpha<2$ is presumed to hold. The inequality $0<\alpha <2$ imposed on the exponent $\alpha$ assures us of the required convergence of the integral
\begin{equation*}
    \int\limits_{\mathbb{R}^n} P(\mathbf{r})\,d\mathbf{r} = 1
\end{equation*}
and meets the divergence of the second moment
\begin{equation*}
    \int\limits_{\mathbb{R}^n} \mathbf{r}^2 P(\mathbf{r})\,d\mathbf{r} = \infty
\end{equation*}
being the characteristic feature of the systems at hand. The purpose of the present Appendix is to relate the given asymptotics of the probability density to corresponding asymptotics of the generating function $G(\mathbf{k})$ as $\mathbf{k}\to0$.

By virtue of \eqref{app.0} and \eqref{appb:1} we have
\begin{multline}\label{appb:2}
    1- G(\mathbf{k}) =   \int\limits_{\mathbb{R}^n} \left(1-
e^{i\mathbf{kr}}\right) P(\mathbf{r})\,d\mathbf{r}
\\
    \rightarrow    2\,\varpi \int\limits_{\mathbb{R}^n}
\sin^2\left(\frac{\mathbf{kr}}{2}\right) \frac1{r^{n+\alpha}}\,d\mathbf{r}
    \quad \text{as}\quad \mathbf{k} \to 0
\end{multline}
because due to the divergence of second moments the region with $|\mathbf{r}|\sim 1/|\mathbf{k}|$ contributes mainly to integral~\eqref{appb:2}.

A direct calculation of the last integral gives rise to the following asymptotic series of the generating function
\begin{equation}\label{appb:3}
    G(\mathbf{k}) = 1 - \omega |\mathbf{k}|^\alpha +
o\left(|\mathbf{k}|^\alpha\right)\,,
\end{equation}
where the coefficient $\omega$ is related to the parameter $\varpi$ as
\begin{equation}\label{appb:4}
    \frac{\varpi}{\omega} = \frac{2^\alpha}{\pi^{\tfrac{n}2+1}}
\sin\left(\frac{\pi \alpha}{2} \right)
    \Gamma\left(1+\frac{\alpha}{2} \right) \Gamma\left(\frac{n + \alpha}{2}
\right)
    \,.
\end{equation}

Dealing with a multidimensional space ($n>1$) we can consider also the random variable $r = |\mathbf{r}|$ whose probability density $P_n(r)$ is related to the probability density $P(\mathbf{r})$ of the random vector $\mathbf{r}$ as
\begin{equation}\label{appb:5}
    P_n(r) = r^{n-1} \oint\limits_{\mathbb{S}_{n,r}} P(\mathbf{r})\,d\sigma
    \,,
\end{equation}
where the surface integral runs over the sphere $\mathbb{S}_{n,r}= \left\{\mathbf{r}: |\mathbf{r}| = r\right\}$. In particular, for the probability density $P(\mathbf{r})$ with the asymptotic behavior \eqref{appb:1} the asymptotic series of probability density $P_n(r)$ is
\begin{equation}
\label{appb:6}
    P_n(r)  =   \frac{\varpi_n}{r^{1+\alpha}} +
o\left(\frac1{r^{1+\alpha}}\right),
\end{equation}
where the coefficient $\varpi_n = S_n \varpi$ and the cofactor $S_n$ is given by expression~\eqref{app.symmS}. Using \eqref{appb:4} we can also write
\begin{equation}\label{appb:7}
    \frac{\varpi_n}{\omega} = \frac{2^{\alpha+1}}{\pi} \sin\left(\frac{\pi
\alpha}{2} \right)
    \Gamma\left(1+\frac{\alpha}{2} \right)
    \frac{\Gamma\left(\dfrac{n + \alpha}{2} \right)}{\Gamma\left(\dfrac{n}{2}
\right)}
    \,.
\end{equation}

The obtained expressions can be interpreted in a reverse manner. Namely, if the generating function $G(\mathbf{k})$ of the random vector $\mathbf{r}$ exhibits the asymptotic behavior~\eqref{appb:3}, then the probability density functions $P(\mathbf{r})$ and $P_n(r)$ of the random vector $\mathbf{r}$ itself and the derived random variable $r = |\mathbf{r}|$, respectively, admit asymptotics~\eqref{appb:1} and \eqref{appb:6} with the corresponding coefficients $\varpi$ and $\varpi_n$ related to the parameter $\omega$ by formulae~\eqref{appb:4} and \eqref{appb:7}.

Let us now measure the spatial scales of $\mathbb{R}^n$ in units of the geometric mean $\overline{r}_g$ of the random vector $\mathbf{r}$ considered in Appendix~\ref{app1} and assume the generating function $G(\mathbf{r})$ to admit the asymptotic behavior~\eqref{appb:3}. Then, according to the obtained results, the random variable $\rho = r/\overline{r}_g$ is characterized by the probability density $\mathcal{P}(\rho)$ with the asymptotics
\begin{equation}\label{appb:f1}
    \mathcal{P}(\rho) = \frac{\varpi_\rho}{\rho^{1+\alpha}}\quad\text{as}\quad
\rho\to \infty\,,
\end{equation}
where $\varpi_\rho = \varpi_n/\overline{r}_g^\alpha$ and, thus,
\begin{multline}\label{appb:f2}
    \varpi_\rho =
    \frac{2^{\alpha+1}}{\pi} \sin\left(\frac{\pi \alpha}{2} \right)
    \Gamma\left(1+\frac{\alpha}{2} \right)
    \frac{\Gamma\Big(\dfrac{n + \alpha}{2} \Big)}{\Gamma\left(\dfrac{n}{2}
\right)}
\\
{} \times \exp\left\{
    \gamma(\alpha-1) + \frac{\alpha}2\left[\Psi\!\Big(\frac{1}{2}\Big) -
\Psi\!\Big(\frac{n}{2}\Big)\right]
    \right\}
\end{multline}
by virtue of \eqref{app.6} and \eqref{appb:7}. Exactly this expression was used in writing expression~\eqref{sec4:4}.

\end{document}